\documentclass[final,twocolumn]{elsart5p}
\journal{Applied Mathematics and Computation}
\usepackage{amsmath,epsfig,cite}

\newcommand{\rmi}{{\rm i}}

\begin{document}
\begin{frontmatter}
\title{Mapped Chebyshev pseudospectral method to study multiple scale phenomena}

\author[uclm]{Adrian Alexandrescu}
\ead{adrian.alexandrescu@uclm.es}
\author[uclm]{Alfonso Bueno-Orovio} 
\ead{alfonso.bueno@uclm.es}
\author[uvigo]{Jos\'e R. Salgueiro}
\ead{jrs@uvigo.es}
\author[uclm]{V\'\i ctor M. P\' erez-Garc\'\i a}
\ead{victor.perezgarcia@uclm.es} 
\address[uclm]{Departamento de Matem\'aticas, E.T.S. Ingenieros Industriales and Instituto de Matem\'atica Aplicada a la Ciencia y la Ingenier\'{\i}a (IMACI), Universidad de Castilla-La Mancha, Avda. Camilo Jos\'e Cela 3, Ciudad Real 13071, Spain}
\address[uvigo]{\'Area de \'Optica, Facultade de Ciencias de Ourense, Universidade de Vigo, As Lagoas s/n, Ourense 32005, Spain}

\begin{abstract}
In the framework of mapped pseudospectral methods, we use a polynomial-type mapping function in order to describe accurately the dynamics of systems developing small size structures. Using error criteria related to the spectral interpolation error, the polynomial-type mapping is compared against previously proposed mappings for the study of collapse and shock wave phenomena.
As a physical application, we study the dynamics of two coupled beams, described by coupled nonlinear Schr\"odinger equations and modeling beam propagation in an atomic coherent media, whose spatial sizes differ up to several orders of magnitude.
It is demonstrated, also by numerical simulations, that the accuracy properties of the polynomial-type mapping outperform in orders of magnitude the ones of the other studied mapping functions.
\end{abstract}

\end{frontmatter}

\section{Introduction}
\label{sec:introduction}

The numerical simulation of physical systems which may develop multiple scale phenomena, like  damage fracture, tumor growth, transport and flow in heterogeneous media, propagation of (non)linear waves, has to be handled with care in order to properly reproduce all their physical features. In such situations the size of the spatial grid and the time advancing step may become critical issues for capturing the dynamics of this type of systems. The numerical difficulties related to a naive increasing of the number of discretization points can be overcome by using more sophisticated techniques, for instance, domain decomposition \cite{Quarteroni1999}, multi-scale finite element method (FEMs) \cite{Brenner2007} or transformations through changes of variables \cite{Boyd1989}. Domain decomposition split the original domain into smaller subdomains which are independently discretized but still linked together by their boundary conditions, which have to ensure a sufficiently smooth solution across the non-matching grids of the different subdomains. Multi-scale FEMs take advantage of the construction of a specific set of basis functions according to the spatial size of each element of the mesh. There is in fact a broad class of FEMs dedicated to the analysis of multiple scale phenomena, each method being designed to address a specific issue, for example, one can capture the large scale behavior of the solution without resolving all the small scale features \cite{Hou1997}. On the other hand, domain transformation methods (or mapping functions) make use of bijective applications to map the points of the physical domain into a computational domain where the function to be discretized is to show a much smoother behavior.

The use of spectral methods has become popular in the last decades for the numerical solution of partial differential equations (PDEs) with smooth behavior due to their increased accuracy when compared to finite-differences or finite-elements stencils with the same degree of freedoms. This happens because the rate of convergence of spectral approximations depends only on the smoothness of the solution, a property known in the literature as ``spectral accuracy''. On the contrary, the numerical convergence of finite-differences or FEMs is proportional to some fixed negative power of $N$, the number of grid points being used. 

For problems with a less smoother behavior, such as those exhibiting rapidly varying solutions, there is a great deal of computational evidence that appropriately chosen mapping functions can significantly enhance the accuracy of pseudospectral applications in thse situations, thus avoiding the use of fine grids and their associated spurious consequences.
Examples include mappings to enhance the accuracy of approximations to shock like functions \cite{Bayliss1987,Bayliss1989,Bayliss1990,Guillard1988,Solomonoff1989,Kosloff1993}, approximation of boundary layer flows in Navier-Stokes calculations \cite{Canuto}, multidomain simulation of the Maxwell's equations \cite{Hesthaven1999}, or cardiac tissue simulations \cite{Zhan2000}. There is also considerable computational evidence that the changes in the differential operator introduced by the mapping do not negatively affect the conditioning of the matrices obtained from the pseudospectral approximation \cite{Guillard1988,Bayliss1987,Bayliss1989,Bayliss1990,Don1997}.

In this work we use a two-parameter polynomial-type mapping function in order to simulate the propagation of two coupled electromagnetic beams of transverse widths as disparate as up to three orders of magnitude. The parameters of the mapping function are adjusted in order to minimize functionals related to the spectral interpolation error. The polynomial mapping is compared against two previously proposed mappings for shock-like fronts and wave collapse phenomena \cite{Bayliss1992,Tee2006}.

The paper is organized as follows. In \mbox{Section \ref{sec:physical_system}} we give a brief description of the underlying physical system. In \mbox{Section \ref{sec:mapping_functions}} the polynomial mapping together with the other mappings are compared using error criteria, and the differences between them are pointed out. In \mbox{Section \ref{sec:numerical_simulations}} the numerical scheme is presented and simulations of the physical system are performed using each mapping. Finally, \mbox{Section \ref{sec:conclusions}} briefly summarizes our main conclusions.

\section{Physical system}
\label{sec:physical_system}

Atomic coherent media were brought into the focus of the scientific community with the theoretical proposal and experimental demonstration of electromagnetic induced transparency (EIT) \cite{Harris1997}. EIT phenomena consists in rendering transparent a rather opaque media by means of an external electromagnetic field, and it is the result of destructive interference between two transition paths having the same final state \cite{Harris1997}. The atomic coherent media exhibits far more physical phenomena \cite{Scully1997}, like lasing without inversion, huge enhancement of refractive index, or negative refractive index \cite{Kastel2007}.   

\begin{figure}
\includegraphics*[angle=-90, width=\columnwidth]{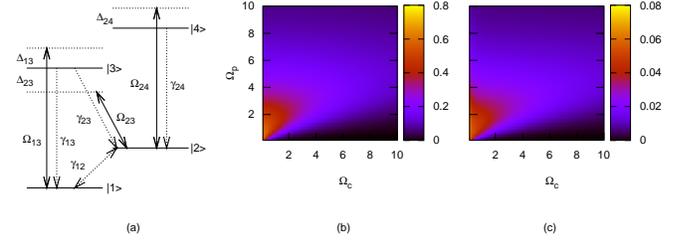}
\caption{Diagram of the energy levels of the atomic system (a), and  
the dependence of nonlinear susceptibilities experienced by the probe, $\chi_p$, (b) and coupling, $\chi_c$, (c) fields.}
\label{fig:physical_model}
\end{figure}

The atomic coherent media of our interest is modeled by a noninteracting atomic gas possesing the four-level energy diagram shown in \mbox{Fig.\ \ref{fig:physical_model}a}.  The atom-fields interaction includes the following parameters: relaxation rates $\gamma_{13}$, $\gamma_{23}$, $\gamma_{24}$, the decoherence rate $\gamma_{12}$ between levels $|1\rangle$ and $|2\rangle$, the amplitudes of electromagnetic fields $\Omega_{13}$, $\Omega_{23}$, $\Omega_{24}$, and the detunings $\Delta_{13}$, $\Delta_{23}$, $\Delta_{24}$ of the field frequency with respect to the energy levels of the atomic media.  A more detailed presentation of our four-level system can be found in Ref. \cite{Hong2002} and the references therein.

Assuming an instantaneous response of the atomic media to the electromagnetic fields, the beams propagation is modeled by a system of two coupled, two-dimensional nonlinear Schr\"odinger (NLS) equations
\begin{subequations}
\begin{eqnarray}
\rmi\frac{\partial\Omega_p}{\partial t} &=& -\Delta \Omega_p - \chi_{p}(|\Omega_p|^2,|\Omega_c|^2)\Omega_p \label{probe_eq}\\
\rmi\frac{\partial\Omega_c}{\partial t} &=& -\Delta \Omega_c - \chi_{c}(|\Omega_p|^2,|\Omega_c|^2)\Omega_c \label{coupling_eq},
\end{eqnarray}
\label{wave_equation_1}
\end{subequations}
where $\Omega_p$ and $\Omega_c$ are respectively known as the probe and coupling (control) fields, and $\chi_p$ and $\chi_c$ are the nonlinear susceptibilities of the atomic media experienced by these probe and coupling fields, respectively. In general, these susceptibilities exhibit both real and imaginary parts. For simplicity, in the present work we neglect the imaginary parts, which are actually associated with the fields absorption. The susceptibilities can then be written in analytical form as the quotient of two bilinear forms of arguments $|\Omega_p|^2$ and $|\Omega_c|^2$, and are similar in structure to those derived in Ref. \cite{Szymanowski1994}:
\begin{equation}
{\chi}_{p,c}=\frac{\sum_{i,j} a_{i,j}^{(p,c)} |\Omega_p|^{2i} |\Omega_c|^{2j}}{\sum_{i,j} b_{i,j} |\Omega_p|^{2i} |\Omega_c|^{2j}}=
\frac{\overline{\Omega}_p^{\scriptscriptstyle{T}} \cdot {\bf A}^{(p,c)} \cdot \overline{\Omega}_c}
{\overline{\Omega}_p^{\scriptscriptstyle{T}} \cdot {\bf B} \cdot \overline{\Omega}_c},
\end{equation}
where $\overline{\Omega}_{p,c}^T=[1\; |\Omega_{p,c}|^{2}\; |\Omega_{p,c}|^{4}\; |\Omega_{p,c}|^{6} \dots |\Omega_{p,c}|^{2m_{p,c}}]$, with
$m_p=6$ and $m_c=5$, are vectors, and ${\bf A}^{(p,c)}=\{a_{i,j}^{(p,c)}\}$, ${\bf B}=\{b_{i,j}\}$ are $(m_p+1) \times (m_c+1)$ matrices. The coefficients of these matrices are sensitive to the values of the fields detunings $\Delta_{13}$, $\Delta_{23}$ and $\Delta_{24}$. For our particular configuration of fields detunings ($\gamma_{12}=10^{-8}\gamma$, $\gamma_{13}=\gamma_{23}=0.6\gamma$, $\gamma_{24}=1.25\gamma$ and $\Delta_{13}=\Delta_{23}=\Delta_{24}=5\gamma$, where $\gamma=30{\rm MHz}$ is a normalization constant) matrices ${\bf A}^p$, ${\bf A}^c$, and {\bf B} are given below. This configuration of detunings was motivated by the cubic-quintic-like model of the NLS equation, which can display liquid light behavior \cite{Michinel2002, Michinel2006}. In \mbox{Fig. \ref{fig:physical_model}b-c} we plot the dependence of the real part of the probe and coupling susceptibilities. 

\begin{subequations}
\begin{eqnarray}
{\bf A}^{p} &=& \left( \begin{tabular}{cccccc} 
0		&0		&0		&0		&0		&0 \\
0   		&0   		&82.707   	&3.323   	&2.1706   	&0 \\
0   		&337.61   	&20.222   	&17.550   	&0   		&0 \\
2.4440   	&19.951   	&35.469   	&0  		&0   		&0 \\
0   		&0   		&0   		&0   		&0   		&0 \\
0   		&0   		&0   		&0   		&0   		&0 \\
0   		&0   		&0   		&0   		&0  		&0 \\
\end{tabular}\right) \times 10^{-5} \\
{\bf A}^{c} &=& \left( \begin{tabular}{cccccc} 
0   &0   	&0		&0		&0   &0 \\
0   &0   	&0		&0		&0   &0 \\
0   &44.799  &1.8004   	&1.1757   	&0   &0 \\
0   &3.4609  &4.7051   	&0		&0   &0 \\
0   &0   	&0		&0		&0   &0 \\
0   &0   	&0		&0		&0   &0 \\
0   &0   	&0		&0		&0   &0 
\end{tabular} \right) \times 10^{-5} \\
{\bf B} &=& \left( \begin{tabular}{cccccc} 
0   		&0		&0   		&209.99	&8.4395   	&5.5115 \\
0		&0   		&1049.7   	&59.073   	&49.605   	&0 \\
0   	     	&1713.6   	&186.72   	&157.28   	&1.3385   	&0 \\
12.411   	&236.32   	&189.51   	&8.9020  	&0   		&0 \\
0.9776   	&7.9273  	&14.187   	&0   		&0   		&0 \\
0   		&0   		&0   		&0   		&0   		&0 \\
0   		&0   		&0   		&0   		&0   		&0 \\
\end{tabular} \right) \times 10^{-5} \nonumber \\
\end{eqnarray}
\end{subequations}

\begin{figure}
\begin{center}
\includegraphics[angle=-90, width=0.85\columnwidth]{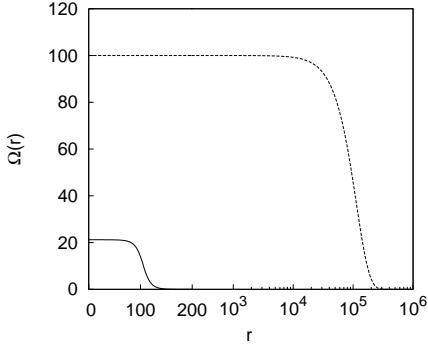}
\caption{The radial profile of the probe (solid line) and of the coupling (dashed line) beam. We remark the different scales of the beam sizes. Starting at $r=200$ the horizontal axis is given in logarithmic scale}
\label{fig:original_beams}
\end{center}
\end{figure} 

In experiments, the spatial transverse width of the coupling beam is much larger that the one of  the probe beam. Therefore, we will study the dynamics of the initial configuration shown in \mbox{Fig.\ \ref{fig:original_beams}}. The coupling field is approximated by a Gaussian function of the form $f(r) = A_0\exp[-(r/w)^{2}]$, with maximum amplitude $A_0 = 100$ and a transverse width $w = 8 \cdot 10^4$. 
Once the control field is properly defined, the probe beam from \mbox{Fig.\ \ref{fig:original_beams}} is computed as a stationary state of \mbox{Eq.\ (\ref{probe_eq})} using a standard shooting method, assuming a spatially constant coupling field $\Omega_c = 100$ in the vicinity of the origin.

\section{Mapping functions}
\label{sec:mapping_functions}
 
Due to their high accuracy and facility to accommodate mapping functions, we choose to discretize the spatial coordinates using a Chebyshev pseudospectral method. In order to properly implement such a method, our infinite domain of interest is first truncated (in each spatial direction) to the interval $[-L,L]$, $L = 5 \cdot 10^5$, and then scaled (without loss of generality) to the interval [-1,1]. This scaling of domains allows the direct use of the Gauss--Lobatto points given by
\begin{equation}
x_j = \cos \left( \frac{\pi j}{N} \right), \quad j = 0, \ldots, N.
\label{eq:Gauss_Lobatto_points}
\end{equation}

A mapping function $g$ is defined as
\begin{equation}
x = g(s,\alpha),
\label{eq:general_mapping}
\end{equation}
where $x$ represents the physical coordinate, $-1 \le s \le 1$ is the computational coordinate (discretized by the Gauss--Lobatto points), and $\alpha$ denotes one or possibly more free parameters. These new sets of collocation points $s$ generated through mappings of the Chebyshev points retain the ease of evaluation of the successive derivatives of a given function. For instance, the first and second derivatives of $u(x)$ can be straightforwardly evaluated as
\begin{subequations}
\begin{eqnarray}
\frac{du}{dx} &=& \frac{1}{g'(s,\alpha)} \frac{du}{ds}, \label{eq:mapped_first_derivative} \\
\frac{d^2u}{dx^2} &=& \frac{1}{[g'(s,\alpha)]^2} \frac{d^2u}{ds^2} - \frac{g''(s,\alpha)}{[g'(s,\alpha)]^3} \frac{du}{ds}, \label{eq:mapped_second_derivative}
\end{eqnarray}
\end{subequations}
For more information related to the use of mappings functions, we refer the reader to Ref. \cite{Boyd1989}.

The profile of our narrow probe beam (see \mbox{Fig. \ref{fig:original_beams}}) exhibits an almost flat region around $x = 0$ before starting its decay to zero. We would like to have its whole support properly discretized, if possible with an almost uniform distribution of points in order to capture all the possible dynamics that might take place along its spatial extent.
 To this intent, we introduce the following polynomial mapping
\begin{equation}
x = (as + s^{2p+1})/(1+a) ,
\label{eq:polynomial_mapping}
\end{equation}
where $a, p > 0$. Adjusting the parameters $a$ and $p$ one can control the size of the region of uniformly distributed points and the number of points located in this region. An almost uniform distributed points near the origin is achieved due to the nonvanishing first derivative of the mapping function $g'(0,\alpha) = a/(1+a)$. Hence, the choice of the parameters $a$ and $p$ have to ensure that, near the origin, the dominant contribution comes from the first order term. Polynomial mappings similar to (\ref{eq:polynomial_mapping}) were used in compresible mixed layer computation \cite{Guillard1992} in order to compare several error functionals of an adaptive pseudospectral method.

\begin{figure}
\includegraphics[angle=-90, width=\columnwidth]{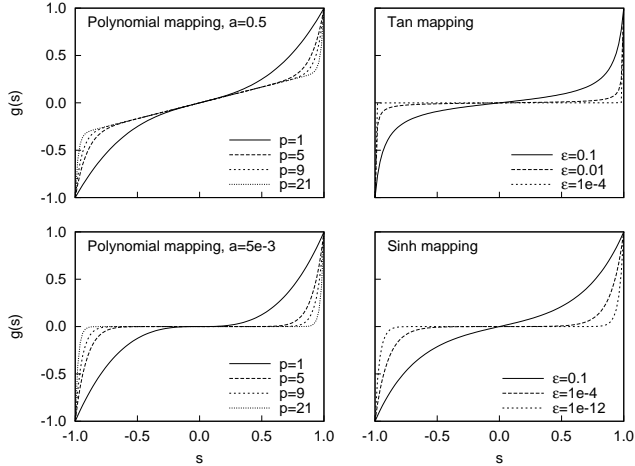}
\caption{Left column: polynomial mapping \eqref{eq:polynomial_mapping} for different values of slope parameter $a$ and polynomial order $p$. Right column: \mbox{``tan-''} and \mbox{``sinh-''} mappings \eqref{eq:tan_mapping}--\eqref{eq:sinh_mapping} for different values of control parameter $\varepsilon$.}
\label{fig:mappings}
\end{figure} 

\begin{figure}
\includegraphics[angle=-90, width=\columnwidth]{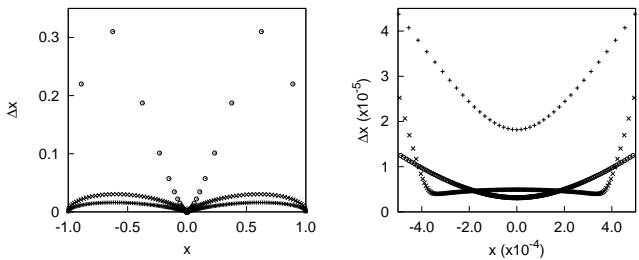}
\caption{Size of the spatial discretization step for the polynomial $(\times)$, \mbox{\mbox{``sinh-''} $(+)$}, and \mbox{\mbox{``tan-''} $(\circ)$} mappings: (left) entire computational domain; (right) central part of comparable size with the narrow probe beam. The mapping parameters used to generate the grids are $a=5.5 \cdot 10^{-4}$, $p=15$ for the polynomial mapping, $\varepsilon = 7.3 \cdot 10^{-5}$ for the sinh-mapping, and $\varepsilon = 2.2 \cdot 10^{-4}$ for the tan-mapping. $N=351$ in all situations.}
\label{fig:step_size}
\end{figure}

We will compare the polynomial mapping against two previously proposed families of mapping functions which also allow a concentration of collocation points in the center of the domain. These mapping functions are given by
\begin{eqnarray}
x & = & \varepsilon \tan( s \tan^{-1} (1/\varepsilon)) \label{eq:tan_mapping} ,\\
x & = & \varepsilon \sinh( s \sinh^{-1} (1/\varepsilon)) \label{eq:sinh_mapping}, 
\end{eqnarray}
where $\varepsilon > 0$. The mapping \eqref{eq:tan_mapping} was introduced in Ref. \cite{Bayliss1992}, and constructed in such a way so that the images of near step functions are almost linear. The mapping \eqref{eq:sinh_mapping} has been recently proposed \cite{Tee2006} for the study of shock waves and blow-up phenomena.
To get more insight into the properties of the mapping \eqref{eq:polynomial_mapping}-\eqref{eq:sinh_mapping}, we plot them and their spatial step size along the whole computational domain, see \mbox{Fig. \ref{fig:mappings}} and \mbox{Fig. \ref{fig:step_size}}, respectively. Optimal parameters are chosen for all mappings as it will be discussed below. It can be observed that both the \mbox{``tan-''} and \mbox{``sinh-''}mappings produce nonuniform step sizes close to $x = 0$, whereas the polynomial mapping is able to produce a discretization grid of almost constant step size in the whole central region.

\subsection{Selection of mapping parameters}
\label{sec:selection_mapping_parameters}

The aim of quantitatively assessing the usefulness of a certain mapping applied to a particular problem has been widely addressed in the literature \cite{Bayliss1987,Guillard1988,Bayliss1989,Bayliss1992}. We follow here the procedure presented in Ref. \cite{Bayliss1992}. Mappings \eqref{eq:polynomial_mapping}--\eqref{eq:sinh_mapping} are functions of one or two parameters which are to be determined. As criteria we will use the functional $I_2$ \cite{Bayliss1992,Guillard1992}, and the $L_2$ and $L_\infty$ norms of the error
\begin{subequations}
\begin{eqnarray}
I_2 & = & \left[ \int_{-1}^1 \frac{(\mathcal{L}^2 f)^2}{\sqrt{1 - s^2}} ds \right]^{1/2}, \label{eq:I2} \\
L_2 & = & \left[ \int_{-1}^1 | f_N (s) -f_M (s) |^2 ds \right] ^{1/2}, \label{eq:L2} \\
L_\infty & = & \max_{s \in [-1,1]} | f_N (s) -f_M (s) |, \label{eq:Linfty}
\end{eqnarray}
\end{subequations}
where $\mathcal{L} = \sqrt{1-s^2} d/ds$. The functional $I_2$ represents an upper bound of the error made when a function is approximated using the first $N$ terms of its Chebyshev expansion \cite{Guillard1992}. The quantity $I_2$ offers a mapping independent criteria. The formulas \eqref{eq:L2} and \eqref{eq:Linfty} compare the $N$ points polynomial interpolation of the function $f$ against the $M$ points one on a larger grid of points, i.e., $N < M$, hence being the $M$ points polynomial interpolation taken as the ``exact'' reference. All integrals are computed using Gauss-Lobatto quadrature formulas. Optimal values for mapping parameters are then selected in order to minimize the above mentioned quantities. 

Our test cases will be conducted in one dimensional space. Nevertheless, as our two dimensional mesh is just the tensor product of the one dimensional grid, the conclusions from the one dimension problem can be straightforwardly extended to the 2D configuration. The top-flat profiles found in the cubic-quintic NLS model are very well approximated by supergaussian functions of the form $f(r) = A_0\exp[-(r/w)^{2m}]$ \cite{Teixeiro1998}. The narrow probe beam profile depicted in \mbox{Fig.\ \ref{fig:original_beams}} can therefore be correctly fitted to this type of profiles, with fitting parameters $A_0\simeq 21.198$, $w\simeq 1.099\cdot10^{2}$, and $m\simeq 4.545$. We will hence use this supergaussian profile as our test/input function. 

\begin{figure}
\includegraphics[angle=-90, width=\columnwidth]{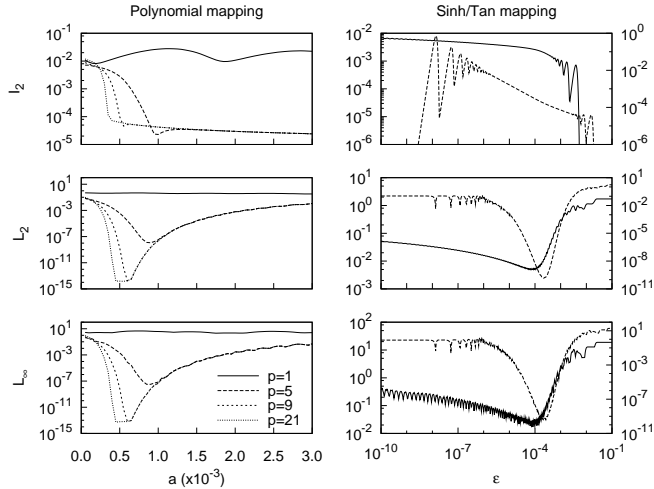}
\caption{Errors in the approximation of the supergaussian profile with the different mappings. Left column: errors using the polynomial mapping \eqref{eq:polynomial_mapping}. Right column: errors using the \mbox{``sinh-''} (solid line) and \mbox{``tan-''} (dashed line) mappings \eqref{eq:tan_mapping}--\eqref{eq:sinh_mapping}. $N=351$ and $M=851$ in all situations. Note the presence of two different scales for the figures on the right column, the left for the \mbox{``sinh-''} and the other for the \mbox{``tan-''} mappings.}
\label{fig:mapping_errors_probe}
\end{figure}

As shown in \mbox{Fig.\ \ref{fig:mapping_errors_probe}} for a number of discretization points $N=351$, the quantities defined by relations \eqref{eq:I2}--\eqref{eq:Linfty} are computed as functions of the different mapping parameters. It was found that, in general, a good mapping will minimize both $I_2$ and $L_2$ quantities at the same time \cite{Bayliss1992}. Optimal values of the mapping parameters were then chosen to minimize the $L_2$ norm of the approximation error, but always comparing the shape of this functional to the ones of $I_2$ and $L_\infty$ in order to ensure that these functionals also attain close to minima values. This choice of criteria was motivated for the unsatisfactory behavior of the functional $I_2$ for the \mbox{``sinh-''} and \mbox{``tan-''} mappings for small values of parameter $\varepsilon$ (due to a poor discretization of the supergaussian profile), as well as for the infinite value of the derivatives of the \mbox{``tan-''} mapping at $x = \pm 1$ as $\varepsilon \rightarrow 0$ (see \mbox{Fig.\ \ref{fig:mappings}}). In addition, the $L_\infty$ functional exhibits in some situations a much bigger oscillatory behavior than the $L_2$ norm, which also makes its use more difficult for the proper choice of the ``optimal parameters''.

Optimal parameters for the correct discretization of the probe field, together with the corresponding values of criteria functions \eqref{eq:I2}--\eqref{eq:Linfty}, are given in \mbox{Table \ref{table:probe}} for the different mappings under study and for two distinct numbers of discretization points, $N = 121$ and $N = 351$. The standard unmapped Chebyshev method is also included for completeness. In the case of $N=121$, the functions $I_2$, $L_2$ and $L_\infty$ exhibit similar shapes to those shown in \mbox{Fig.\ \ref{fig:mapping_errors_probe}}, but with sharper minima due to the smaller number of sample points.  In all situations, our polynomial mapping is found to outperform the results obtained using the other mapping functions due to its ability of generating an almost uniform discretization grid in the whole extent of the narrow beam. In addition, it is noteworthy to remark that the values of optimal parameters $a$ and $p$ are noncritical. Similar results are obtained when compared to other mappings found in the literature, such as those described in \cite{Boyd1992, Kosloff1993}.

\begin{table}
\caption{Error comparison for the probe field when using the polynomial \eqref{eq:polynomial_mapping}, \mbox{``tan-''} \eqref{eq:tan_mapping} and \mbox{``sinh-''} \eqref{eq:sinh_mapping} mappings. $M = 851$ in all situations. U denotes unmapped.}
\begin{center}
\begin{tabular}{ccccc}
\shortstack[c]{Mapping\\ $ $}&\shortstack[c]{Optimal parameters\\($N=121$)}&\shortstack[c]{$I_2$\\$ $}&\shortstack[c]{$L_2$\\$ $}& \shortstack[c]{$L_\infty$\\$ $} \\
\hline
\eqref{eq:polynomial_mapping} & $a\simeq 4$e-04, $p=21$ & 7.1657e-05 & 1.2179e-08 & 3.2862e-08 \\
\hline
\eqref{eq:tan_mapping} & $\varepsilon\simeq 2.1412$e{-04} & 1.7164e-04 & 8.8775e-04 & 1.7989e-03 \\
\hline
\eqref{eq:sinh_mapping} & $\varepsilon\simeq 7.2731$e-05 & 3.1044e-03 & 3.5283e-01 & 1.6369 \\
\hline
U & -- & NA & 3.2627 & 20.918 \\
\hline 
& & & \\
\end{tabular}
\begin{tabular}{ccccc}
\shortstack[c]{Mapping\\ $ $}&\shortstack[c]{Optimal parameters\\($N=351$)}&\shortstack[c]{$I_2$\\$ $}&\shortstack[c]{$L_2$\\$ $}& \shortstack[c]{$L_\infty$\\$ $} \\
\hline
\eqref{eq:polynomial_mapping} & $a\simeq 5.5$e-04, $p=15$ & 5.8775e-05 & 1.3730e-14 & 4.9737e-14 \\
\hline
\eqref{eq:tan_mapping} & $\varepsilon\simeq 2.2320$e-04 & 1.6753e-04 & 1.4671e-10 & 3.6194e-10 \\
\hline
\eqref{eq:sinh_mapping} & $\varepsilon\simeq 7.2731$e-05 & 2.7488e-03 & 4.9130e-03 & 1.6885e-02 \\
\hline
U & -- & NA & 1.7230 & 18.882 \\
\hline
\end{tabular}
\end{center}
\label{table:probe}
\end{table}

From the results presented in \mbox{Table \ref{table:probe}}, it can be inferred that the polynomial mapping \eqref{eq:polynomial_mapping} is much more accurate than the \mbox{``sinh-''} mapping even when using optimal values for parameter $\varepsilon$, because the latter produces much bigger step sizes close to the origin. Furthermore, for the \mbox{``sinh-''} mapping the $I_2$ functional does not seem to behave as an upper bound of the $L_2$ and $L_\infty$ norms, as it was previously demonstrated in \mbox{Ref.\ \cite{Guillard1992}}. This points out a possible poor discretization of the function under representation. In fact, the number of points has to be increased till $N = 551$ in order to have these inequalities satisfied when using this mapping. The same happens when using the \mbox{``tan-''} mapping and a small number of discretization points ($N = 121$). The value of functional $I_2$ is not assigned (NA) for the unmapped Chebyshev method since in this situation the probe field is discretized by a single collocation point. 

However, our system of interest consists in two coupled beams, and therefore the coupling field has to be also properly discretized for our choice of mapping parameters. \mbox{Table \ref{table:coupling}} presents values of functionals \eqref{eq:I2}--\eqref{eq:Linfty} for the coupling field for the choice of parameters that best discretizes the narrow supergaussian profile. Even with a reduced number of collocation points ($N=121$), the polynomial mapping is able to produce a fairly good description of this field, and of comparable accuracy to the best of the other mappings when the spatial resolution is increased ($N=351$). On the other hand, the \mbox{``tan-''} mapping is not capable of describing correctly this wider profile, since it concentrates almost all discretization points in the center of the interval. The \mbox{``sinh-''} mapping, as well as the unmapped Chebyshev method, is able to discretize the control field, but was not able to represent appropriately the narrow probe field.

\begin{table}
\caption{Error comparison for the coupling field when using the polynomial \eqref{eq:polynomial_mapping}, \mbox{``tan-''} \eqref{eq:tan_mapping} and \mbox{``sinh-''} \eqref{eq:sinh_mapping} mappings, using the sets of parameters which give optimal description of the probe field. $M = 851$ in all situations. U denotes unmapped.}
\begin{center}
\begin{tabular}{ccccc}
\shortstack[c]{Mapping\\ $ $}&\shortstack[c]{Optimal parameters\\($N=121$)}&\shortstack[c]{$I_2$\\$ $}&\shortstack[c]{$L_2$\\$ $}& \shortstack[c]{$L_\infty$\\$ $} \\
\hline
\eqref{eq:polynomial_mapping} & $a\simeq 4$e-04, $p=21$ & 5.6487e-03 & 7.5856e-05 & 2.4692e-04 \\
\hline
\eqref{eq:tan_mapping} & $\varepsilon\simeq 2.1412$e{-04} & 5.2026e-02 & 5.8214e-01 & 5.0589 \\
\hline
\eqref{eq:sinh_mapping} & $\varepsilon\simeq 7.2731$e-05 & 1.8944e-03 & 1.4784e-12 & 2.5579e-12 \\
\hline
U & -- & 9.3423e-05 & 2.9361e-14 & 1.2789e-13 \\
\hline 
& & & \\
\end{tabular}
\begin{tabular}{ccccc}
\shortstack[c]{Mapping\\ $ $}&\shortstack[c]{Optimal parameters\\($N=351$)}&\shortstack[c]{$I_2$\\$ $}&\shortstack[c]{$L_2$\\$ $}& \shortstack[c]{$L_\infty$\\$ $} \\
\hline
\eqref{eq:polynomial_mapping} & $a\simeq 5.5$e-04, $p=15$ & 4.6172e-03 & 1.2273e-13 & 3.4106e-13 \\
\hline
\eqref{eq:tan_mapping} & $\varepsilon\simeq 2.2320$e-04 & 4.0986e-02 & 2.6578e-03 & 9.0893e-03 \\
\hline
\eqref{eq:sinh_mapping} & $\varepsilon\simeq 7.2731$e-05 & 1.8995e-03 & 1.1528e-13 & 3.6948e-13 \\
\hline
U & -- & 9.3678e-05 & 4.9873e-14 & 2.4158e-13 \\
\hline
\end{tabular}
\end{center}
\label{table:coupling}
\end{table}

\section{Numerical simulations}
\label{sec:numerical_simulations}
The propagation of the probe and coupling fields is simulated using a split-step mapped pseudospectral method as the one presented in \mbox{Ref.\ \cite{Montesinos2005}}.
The linear step (Laplace operator) is integrated by using exponential integration of the transformed Chebyshev matrix, whereas the nonlinear step is performed by using explicit midstep Euler method. In order to ensure transparent boundary conditions, we have placed an absorbing potential to get rid of the potentially outgoing radiation \cite{Montesinos2005}. Using this numerical scheme we have simulated the time evolution of the initial probe and coupling fields shown in \mbox{Fig.\ \ref{fig:original_beams}}, given by the NLS system \eqref{wave_equation_1}, for all the three mappings given in the previous section. The parameters of the mappings were kept fixed during the time evolution. The time step and the number of sample points are set to $\Delta t=0.1$ and $N=121$, respectively. As the initial fields do not constitute a stationary solution of the coupled NLS system \eqref{wave_equation_1}, they will change their shape in the course of the numerical simulation. 
We have verified that the computational results shown bellow are not altered when changing the size of time step, e.g., $\Delta t=0.01$ or 1.

\begin{figure}
\includegraphics[angle=-90, width=\columnwidth]{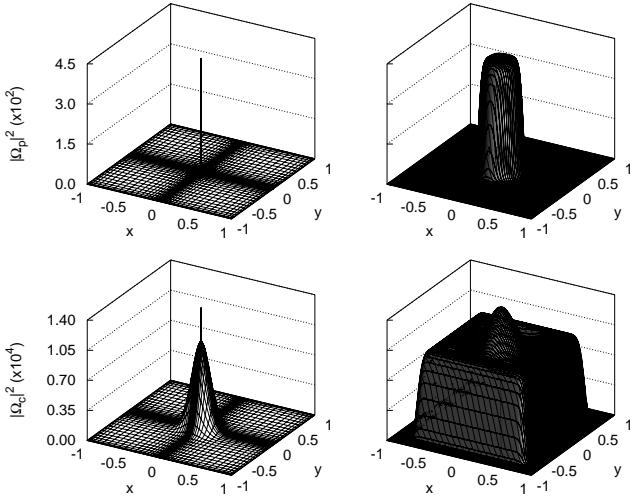}
\caption{Field amplitudes for $t=3600$ computed with the polynomial-mapped Chebyshev grid, with $N=121$, $a=5\cdot 10^{-4}$ and $p=12$. Upper (lower) row shows the probe (coupling) field, while the left (right) column depicts the spatial profiles on physical (computational) domain.}
\label{fig:field_poly_samples_z=3600}
\end{figure}

\begin{figure}
\includegraphics[angle=-90, width=\columnwidth]{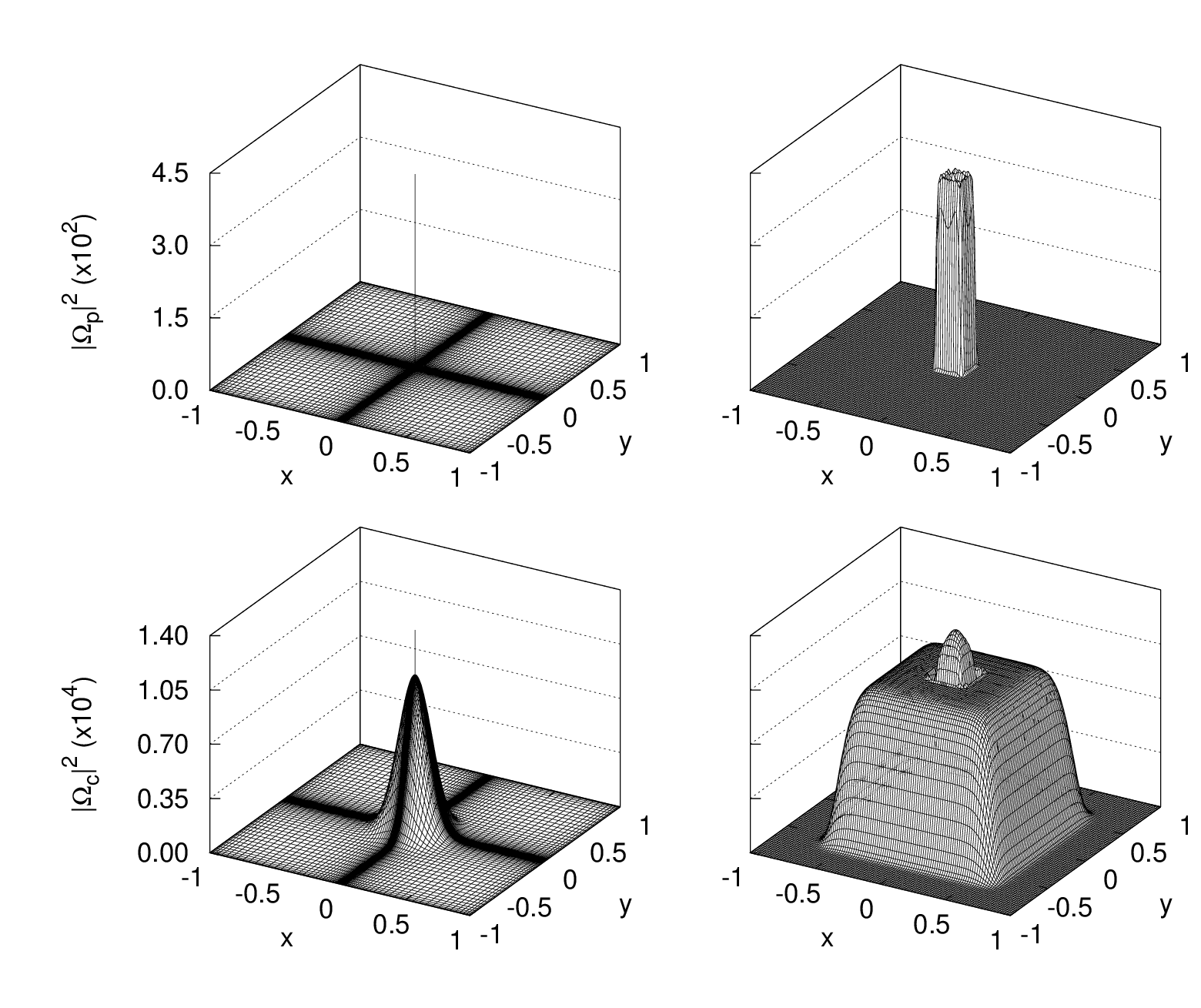}
\caption{Field amplitudes for $t=3600$ computed with the \mbox{``sinh''}-mapped Chebyshev grid, with $N=121$ and $\varepsilon=7.2731\cdot 10^{-5}$. Upper (lower) row shows the probe (coupling) field, while the left (right) column depicts the spatial profiles on physical (computational) domain.}
\label{fig:field_sinh_samples_z=3600}
\end{figure}

\begin{figure}
\includegraphics[angle=-90, width=\columnwidth]{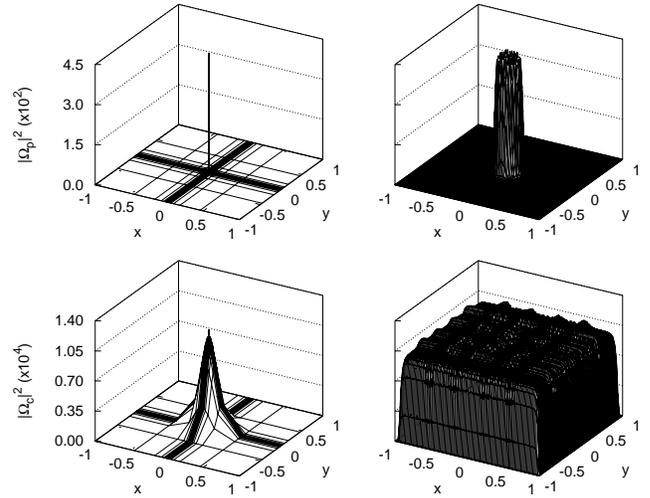}
\caption{Field amplitudes for $t=600$ computed with the \mbox{``tan''}-mapped Chebyshev grid with $N=121$ and $\varepsilon=4.5217\cdot 10^{-4}$. Upper (lower) row shows the probe (coupling) field, while the left (right) column depicts the spatial profiles on physical (computational) domain.}
\label{fig:field_tan_samples_z=600}
\end{figure}

In \mbox{Figs.\ \ref{fig:field_poly_samples_z=3600}-\ref{fig:field_tan_samples_z=600}} we plot the spatial profiles of the probe and coupling fields on both the physical and computational domains. Around $t\simeq 3600$ the dynamics shows the developing of a peak into the coupling beam $\Omega_c$, of comparable spatial width with the narrow probe beam, while the probe field only exhibits slight modifications of its spatial profile. 
In the case of the polynomial-mapped Chebyshev grid, both the probe and coupling fields show smooth variations in the associated computational domain, with their peaks and spatial decays correctly sampled. In especial, note how the almost singular structure that represents the probe field is perfectly approximated by this mapping even using a small number of grid points ($N=121$). On the other hand, the use of the \mbox{``sinh''}-mapped Chebyshev grid leads to a merely rectangular probe profile $\Omega_p$, with a poor sampling of its spatial decay, see the upper-right plot of the \mbox{Fig. \ref{fig:field_sinh_samples_z=3600}}. This fact is also manifested on the peak located in the center of the coupling beam, see the lower-right plot of \mbox{Fig.\ \ref{fig:field_sinh_samples_z=3600}}.

In the case of the \mbox{``tan''}-mapped Chebyshev grid, see \mbox{Fig. \ref{fig:field_tan_samples_z=600}}, due to its poor spatial discretization, the coupling beam is quickly polluted, by $t\simeq 600$, with significant errors. These errors are coupled back into the probe beam which shows a noisy spatial profile. Hence, the subsequent time development of the system is altered.

\section{Conclusions}
\label{sec:conclusions}

In order to study the propagation of two coupled beams exhibiting spatial widths of several orders of magnitude of difference, we have used a two-parameter polynomial-type mapping function especially suitable for its use in conjunction with Chebyshev pseudospectral methods. Using error criteria related to the spectral accuracy, we have compared the approximation error attained by the polynomial-type mapping against the ones obtained using previously defined mappings proposed to capture collapse or shock wave phenomena. We have also performed numerical simulations of two coupled beams propagating through an atomic coherent media, where the propagation is described by a system of two coupled NLS equations. While the \mbox{``sinh''}-mapping and \mbox{``tan''}-mappings only offer proper discretizations of the coupling and probe beams, respectively, the polynomial-mapping is able to capture simultaneously all the physical features of both fields, still using a relatively small number of discretization points. The results from the comparison of the error criteria presented in \mbox{Section \ref{sec:mapping_functions}} are also supported by numerical simulations. Furthermore, the results presented in \mbox{Fig.\ \ref{fig:mapping_errors_probe}} indicate that the optimal values of the polynomial-mapping parameters are noncritical. 

It is worth emphasizing the easiness of implementation of the proposed mapping in comparison with the implementation of either a multiple scale or domain decomposition method. In addition, a third parameter, corresponding to the center of the uniform discretized region, can be easily accommodated into the polynomial mapping, allowing the tracking of moving and interacting structures of small spatial size.

\section{Acknowledgments}
The authors thank to H. Michinel for initiating the discussion on the realization of light condensates in atomic coherent media, from which the present work has been developed. This work was supported by grants \mbox{FIS2006--04190} (Ministerio de Educaci\'on y Ciencia, Spain), \mbox{PAI-05-001} and \mbox{PCI-08-0093} (Consejer\'ia de Educaci\'on y Ciencia de la Junta de Comunidades de Castilla-La Mancha, Spain). The work of the first author was supported by the Ministerio de Educaci\'on y Ciencia (Spain) under Grant No.\ AP-2004-7043.

\end{document}